\documentclass[fleqn,usenatbib]{mnras}
\usepackage{newtxtext,newtxmath}

\usepackage[T1]{fontenc}
\DeclareRobustCommand{\VAN}[3]{#2}
\let\VANthebibliography\thebibliography
\def\thebibliography{\DeclareRobustCommand{\VAN}[3]{##3}\VANthebibliography}

\usepackage{graphicx}	% Including figure files
\usepackage{amsmath}	% Advanced maths commands
\usepackage{amssymb}	% Extra maths symbols

\title[Standardization of stellar luminosities]{Standard stellar luminosities; what are typical and limiting accuracies in the era after {\it Gaia}?}

\author[Eker et al.]{
Z. Eker,$^{1}$\thanks{E-mail: eker@akdeniz.edu.tr}
F. Soydugan,$^{2,3}$
S. Bilir,$^{4}$
and V. Bak{\i}\c{s}$^{1}$
\\
$^{1}$Akdeniz University, Faculty of Sciences, Department of Space Sciences and 
Technologies, 07058, Antalya, Turkey\\
$^{2}$Department of Physics, Faculty of Arts and Sciences, \c{C}anakkale Onsekiz 
Mart University, 17100 \c{C}anakkale, Turkey\\
$^{3}$Astrophysics Research Center and Ulup{\i}nar Observatory, \c{C}anakkale 
Onsekiz Mart University, 17100, \c{C}anakkale, Turkey\\
$^4$Istanbul University, Faculty of Science, Department of Astronomy and Space
Sciences, 34119, Istanbul, Turkey\\
}
\date{Accepted XXX. Received YYY; in original form ZZZ}

\pubyear{2015}

\begin{document}
\label{firstpage}
\pagerange{\pageref{firstpage}--\pageref{lastpage}}
\maketitle

\begin{abstract}
Methods of obtaining stellar luminosities ($L$) have been revised and a new concept, standard stellar luminosity, has been defined. Among the three methods (direct method from radii and effective temperatures, method using a mass-luminosity relation (MLR) and method requiring a bolometric correction), the third method, which uses the unique bolometric correction ({\it BC}) of a star extracted from a flux ratio ($f_{\rm V}/f_{\rm Bol}$) obtained from the observed spectrum with sufficient spectral coverage and resolution, is estimated to provide an uncertainty ($\Delta L/L$) typically at a low percentage, which could be as accurate as 1\% perhaps more. The typical and limiting uncertainties of the predicted $L$ of the three methods were compared. The secondary methods requiring either a pre-determined non-unique {\it BC} or MLR were found to provide less accurate luminosities than the direct method, which could provide stellar luminosities with a typical accuracy of 8.2\% - 12.2\% while its estimated limiting accuracy is 2.5\%.
\end{abstract}


%%%%%%%%%%%%%%%%%%%%%%%%%%%%%%%%%%%%%%%%%%%%%%%%%%

%%%%%%%%%%%%%%%%% BODY OF PAPER %%%%%%%%%%%%%%%%%%

\section{Introduction}
Luminosity of a star is not a directly observable parameter. It is, rather, an empirical parameter to be computed from observable parameters; a radius ($R$) and an effective temperature ($T_{\rm eff}$). Therefore, the most reliable stellar luminosities so far are the ones which are calculated directly from the Stefan-Boltzmann law ($L=4\pi R^2\sigma T_{\rm eff}^4$) from observed radii and effective temperatures of Detached Double-lined Eclipsing Binaries (DDEB) \citep*{Andersen91, Torresetal10, Eker14, Eker15, Eker18, Eker20} obtained from simultaneous solutions of their radial velocity and eclipsing light curves and/or analysis of disentangled spectra of the components \citep{Hadrava95, Bakis07}. Except for a very limited number of nearby single stars which have radii available by interferometry and any kind of observed effective temperatures, the direct method, alas, has a serious defect that it is not applicable to single stars, visual and spectroscopic binaries (or multiple systems) because their stellar radii are not observable directly. Moreover, effective temperatures implied by intrinsic colors, in most cases, are also not observable due to interstellar reddening. The direct method even faces difficulties for contact, semi-detached and even close binaries because of proximity effects which deform star shapes, thus the Stefan-Boltzmann law is not applicable directly.      

In addition to this limited primary source of stellar luminosities (the direct method), there are two other secondary sources, which indirectly provide stellar luminosities. They are indirect because both rely upon a prior relation determined by the most accurate stellar parameters mostly from DDEB.

\subsection{Mass-Luminosity relations for predicting L}
From the Historical point of view, the first of the secondary sources is the main-sequence mass-luminosity relation ($L \propto M^{\alpha}$). Despite being applicable to only main-sequence stars, this method has the power to enlarge the availability of stellar luminosities towards single stars (if their masses are known or estimated somehow) and visual binaries (or multiple systems) with orbital parameters. If orbital parameters could be extracted from a visual orbit, then component masses would be known according to Kepler's third law. Note that, unlike a visual binary with an orbital inclination deduced from its visual orbit, a spectroscopic binary is without an orbital inclination because it is not possible to deduce the orbital inclination from the observed spectra. Therefore, spectroscopic binaries do not provide the true masses of components except for a mass ratio or a mass function, hence this method is not applicable to spectroscopic binaries unless orbital inclinations are somehow available. 

The main-sequence mass-luminosity relation (MLR) was discovered independently by \citet{Hertzsprung23} and \citet*{Russell23} empirically in the middle of the first half of the 20th century. As newer and more accurate data came along, it has been revised, updated and improved \citep*{Eddington26, McLaughlin27, Kuiper38a, Petrie50a, Petrie50b, Strand54, Eggen56, McCluskey72, Cester83, Griffiths88, Demircan91} many times. Early relations were demonstrated as mass-absolute bolometric magnitude diagrams, some with the best fitting curve and some without. First, \citet{Eggen56} attempted to define the power of mass (alpha) so the relation is expressed as $L=\mu^{3.1}$, where $\mu$ is defined as $\mu=a^3/P^2 \varpi^3$ from Kepler's Harmonic Law in which $a$ and $\varpi$ are semimajor axis and parallax of the double star in units of arcseconds, $P$ is orbital period in years, and $\mu$ mass of the system in solar masses. While \cite{McCluskey72} tried to establish the relation as $M=\alpha L^{\beta}$, where $M$ and $L$ are the masses and luminosities of components in which $\alpha$ and $\beta$ are the constants determined on various mass-absolute bolometric magnitude diagrams, \citet{Cester83}, \citet{Griffiths88} and \citet{Demircan91} preferred to study on a mass-luminosity diagram in order to define unknown constants on the classical form of mass-luminosity ($L\propto M^{\alpha}$) relation either by fitting a curve to all data or dividing the mass range into two (low mass stars, high mass stars) or three (high mass, intermediate mass or solar, low mass) in order to define the inclination of the linear MLR (power of $M$) and its zero point constant on $\log M-\log L$ diagrams.

\citet{Andersen91} objected to defining any form of MLR and preferred to display the $\log M-\log L$ diagram without a curve fit because the scatter from the curve is not only due to observational errors but also due to abundance and evolutionary effects. He claimed: ``... departures from a unique relation are real'', so if there is no unique function to represent data, why bother to define one? Because of this objection, \citet{Henry04} and \citet{Torresetal10} also displayed their diagrams without a curve. Except for \citet{Gorda98}, who preferred the form $M_{\rm Bol} = a + b \log M$, where $M_{\rm Bol}$ is the absolute bolometric magnitude and $M$ is the mass, and \citet{Henry93}, who preferred $\log M = aM_{\xi} + b$ for infrared colours, where $M_{\xi}$ stands for absolute magnitude at the $J$, $H$ and $K$ bands, and $\log M=a M_V^2+b M_V+c$ for the $V$ band to express various MLR with unknown coefficients $a$, $b$, and $c$ to be determined by data on the diagram, and \cite{Malkov07}, who defined MLR and inverse MLR functions, authors such as \citet*{Gafeira12}, \citet{Torresetal10}, \citet{Benedict16}, \citet{Moya18} and recently, \citet*{Fernandes21} deviated from the tradition of defining MLR, which could be used on both directions ($L$ computed from $M$ or $M$ computed from $L$), with the idea that the luminosity of solar type single stars could be obtained from observations with fair accuracy but not the mass \citep{Fernandes21}. Therefore, MLR should be established only for estimating the mass of single stars from other astrophysical stellar parameters such as its luminosity, metallicity ($Z$) and age, not for estimating luminosities from masses \citep{Fernandes21}. Nevertheless, the predicted relation is still called MLR despite it not being a relation solely between mass and luminosity but also including metallicity and age as the observable parameters. Devised for estimating mass rather than luminosity, MLR of this kind is not suitable for this study.   

Classical MLR in the form $L \propto M^{\alpha}$ was appreciated by \citet{Ibanoglu06} when they were comparing mass-luminosity relations for detached and semidetached Algols. The tradition of MLR in the form $L\propto M^{\alpha}$ (or reducible to it) has been continued by \citet{Eker15, Eker18}. Note that any curve or a polynomial of any degree fitting data on a $\log M - \log L$ diagram is reducible to the form $L\propto M^{\alpha}$ because the derivative of the fitting function at a given mass gives the value of alpha. The advantage of such MLR is not only that it works both ways ($L$ from $M$, or $M$ from $L$), but also because it permits one to relate typical masses and luminosities of main sequence stars in general. Since this study is primarily interested in estimating typical accuracies of mass and luminosities, the six-piece classical MLRs of \citet{Eker18}, as the most recent determined MLRs, are more suitable for this study than any of the other MLR forms. 

Although $L$ obtained from a classical MLR would be a unique value for a given mass ($M$), it is akin to the mean value of all luminosities from the Zero Age Main-Sequence (ZAMS) to Terminal Age Main-Sequence (TAMS) of stars with the same mass but with different age and various chemical composition. Thus, the uncertainty of obtained $L$ is expected to be very large for those who are looking for an accurate $L$ of a star in question. Still, this was the only method for producing the luminosities of single stars with known masses and visual binary or multiple systems with visual orbits in earlier times, when the third source using bolometric corrections ({\it BC}) was not available yet or the {\it BC} values were not as accurate as today.

\subsection{Bolometric Corrections for predicting L}

Perhaps the most powerful secondary method for obtaining $L$ is the method using bolometric correction {\it (BC)}. It appears to be even more powerful than the direct method not only because it is applicable to all stars, single or binary (or multiple), but also because it is more practical and easier to use. The method can even work with a single observation at a preferred filter, let us say $V$ filter, which nowadays it could be accurate at a milimag level, perhaps more accurate, if the star is bright enough, though not a binary or a multiple system. For binaries and multiple systems, however, the method requires the relative light contributions of the components. Only disadvantage, compared to the direct method, is that a trigonometric parallax ($\varpi$) and reddening (or extinction) of the stars must be provided. Nevertheless, despite these advantages, the method is still secondary because it does not work if a pre-determined {\it BC} value is not available, which could be read from any of the tabulated {\it BC} tables in the literature, or from $BC-T_{\rm eff}$, $BC-M/M_{\odot}$ \citep{Flower96, Eker20, Eker21}, or similar relations if available. 

Analytical $BC-T_{\rm eff}$ relations, however, are determined only by two authors \citep{Flower96, Eker20}, the rest of the available {\it BCs} are all in tabular format \citep*{Kuiper38b, Popper59, McDonald52, Johnson64, Johnson66, Heintze73, Code76, Malagnini85, Cayrel97, Bessell98, Girardi08, Sung13, Chen19} where the variation of {\it BC} values with the other stellar parameters such as spectral type, intrinsic colour, luminosity class, metallicity and surface gravity may also be given. 

It was \citet{Torres10} who first noticed inconsistencies in the use of $BC_{\rm V}$ values that may lead to errors of up to 10\% or more in the derived luminosity equivalent and about 0.1 mag or more uncertainty in the bolometric magnitudes. According to \citet{Torres10}, the problems arise from the arbitrariness attributed to the zero point of the {\it BC} scale. Recently, \citet{Eker21} revised the zero points of the {\it BC} scales on the tabulated tables and confirmed \citet{Torres10} independently. According to the results of \citet{Eker21} there could be up to 0.1 mag systematic shifts of {\it BC} values corresponding up to 10\% systematic errors in predicted stellar luminosities, which are intolerable in the era after {\it Gaia}. 

In this paper, we must first re-emphasize IAU 2015 General Assembly Resolution B2, which \citet{Eker21} relied upon in solving the problems originating from the arbitrariness attributed to the {\it BC} scale. Fixing the zero point of the {\it BC} according to IAU 2015 General Assembly Resolution would actually mean the standardization of {\it BC}. Briefly, the standardization of {\it BC} values was a solution suggested by \citet{Eker21} to remove uncertainties both on stellar absolute bolometric magnitudes and predicted stellar luminosities caused by the arbitrariness attributed to the zero point constants of {\it BC} values appearing in the literature.  

Since stellar luminosities are used by Galactic and extragalactic astronomers to estimate Galactic and extragalactic structures and luminosities, which could then be used to estimate Galactic and extragalactic distances as well as the luminous mass contained in galaxies and universe, the standardization apparently is expected to have a widespread effect on dark matter research, Hubble law and cosmological models. For this reason, one must be careful about uncertainties and errors reflected in {\it BC} values, which naturally propagate to stellar luminosities. Not only the luminosities of stars, but also for deep space and cosmology researches, are affected. This study intends to explain (1) how these negative propagating effects could be minimized and (2) how to estimate and compare accuracies of the predicted stellar luminosities by the three methods, as well as (3) the need to emphasize the differences between standard and non-standard luminosities. 

\section{Data}

Preliminary data of this study originates with to the ``Catalog of Stellar parameters from the Detached Double-Lined Eclipsing Binaries in the Milky Way'' containing 514 stars (257 systems) by \citet{Eker14}. Although the updated catalog \citep{Eker18} was expanded to include 639 stars (318 binaries and one eclipsing spectroscopic triple), after removing the stars having errors greater than 15\% both on mass and radius, and eliminating the stars belonging to globular clusters, then finally choosing main-sequence stars with a mass of $0.179\leq M/M_{\odot}\leq 31$ and metal abundance $0.008\leq Z\leq 0.040$ ranges within the limits of theoretical ZAMS (Zero Age Main Sequence) and TAMS (Terminal Age Main Sequence) according to PARSEC models \citep{Bressan12}, 509 stars were retained by \citet{Eker18} to study the MLR of the sample representing the nearby stars in the Galactic disc in the solar neighborhood. 

The most accurate stellar luminosities and propagated uncertainties were calculated according to the direct method (Method 1) using the most accurate radii and effective temperatures and the associated observed uncertainties of the 509 main-sequence stars, which are the ``components of DDEB'' in the updated catalog. The six-piece MLRs in the form of $\log L = a \log M + b$ calibrated by the data on the $\log M-\log L$ diagram covering the mass $0.179\leq M/M_{\odot}\leq 31$ and metal abundance $0.008\leq Z\leq 0.040$ ranges for the main-sequence stars in the Galactic disc in the solar neighborhood by \citet{Eker18} are taken for granted in this study.  

\section{Relative accuracies According to three methods}
\subsection{Relative uncertainty of L according to Method 1}
Uncertainty of luminosity by the direct method using the Stefan-Boltzmann law ($L = 4\pi R^2 \sigma T_{\rm eff}^4$) could be calculated by:

\begin{equation}
{\frac{\Delta L}{L}}=\sqrt{\left(2\frac{\Delta R}{R}\right)^2+\left(4\frac{\Delta T_{\rm eff}}{T_{\rm eff}}\right)^2.}
\end{equation}
where, $\Delta R/R$ and $\Delta T_{\rm eff}/T_{\rm eff}$ are relative observational random errors of a star's radius and effective temperature, which usually comes from simultaneous solutions of the light and radial velocity curves of DDEB and spectral analysis of the disentangling spectra of the system's components. 

\subsection{Relative uncertainty of L according to Method 2}
In order to estimate relative uncertainty of $L$ using the method of error propagation, the form of adopted MLRs ($L\propto M^{\alpha}$) indicates
\begin{equation}
{\frac{\Delta L}{L}}=\alpha \frac{\Delta M}{M}
\end{equation}
where $\Delta M/M$ is the relative uncertainty of the observed mass of a star, $\alpha$ is the power of $M$ and $\Delta L/L$ is the relative uncertainty of the predicted luminosity. According to \citet{Eker15, Eker18}, the equation (2) is invalid because the dispersions on the $\log M-\log L$ diagram are not only due to observational uncertainties of $M$, but also due to the age and chemical composition differences of the stars in the sample \citep{Andersen91, Torresetal10, Eker15, Eker18}. It is better to use: 
\begin{equation}
{\frac{\Delta L}{L}}=\frac{SD}{0.4343}
\end{equation}
where $SD$ is the standard deviation of data from the MLR function, which should be chosen according to the mass of the star in question. There are six MLR functions with standard deviations and inclinations, already computed by \citet{Eker18}, which will be used in this study. Only if the cases are $\alpha \frac{\Delta M}{M}>\frac{SD}{0.4343}$ would equation (2) then be valid. Because the typical relative uncertainties of $M$ are in the order of 1-2\% \citep{Eker14}, equation (2) is not valid. It would be valid for stars with relative uncertainties bigger than about 6\%, for low mass stars ($M<2.4M_{\odot}$), and relative uncertainties bigger than about 10\% for high mass stars ($M>2.4M_{\odot}$) \citep{Eker15}.

\subsection{Relative uncertainty of L according to Method 3}
The same data set of 509 main-sequence stars used for method 2 above is also used for method 3 here, which uses a pre-determined bolometric correction ({\it BC}) for predicting the luminosity of a star. Unfortunately, many of the binary systems containing the 509 stars as components of DDEB had to be eliminated because some of the binaries have not been observed in standard $V$ magnitudes or otherwise, the light contributions of components in the $V$ band could not be achieved, or a reliable trigonometric parallax for the star ($\varpi$) did not exist, or reliable interstellar reddening could not be found.

\citet{Eker20} could find only 206 binaries which have at least one component in the main-sequence (194 systems with components, eight systems with primaries, four systems with secondaries on the main-sequence), leaving a total of 400 main-sequence stars which are eligible to compute bolometric correction ({\it BC}) coefficients in the $V$ band. A standard $BC_{\rm V}-T_{\rm eff}$ curve (a fourth degree polynomial) which is valid in the range $3100\leq T_{\rm eff}\leq 36000$ K was calibrated and the coefficients of the polynomial, errors of the coefficients, standard deviation ($SD = 0.215$) and correlation coefficient ($R^2 = 0.941$) have already been announced by \citet{Eker20}. Containing the necessary statistics Table 5 of \citet{Eker20} was adopted for this study as the basic data to compute the luminosity of a star ($L$) according to Method 3.      

In this method, the $L$ of a star could be achieved according to the following relation:  
\begin{equation}
M_{\rm Bol}=-2.5\log L + C_{\rm Bol},
\end{equation}
where $M_{\rm Bol}$ is the absolute bolometric magnitude of the star and $C_{\rm Bol}=71.197~425~...$ or if $L$ is in SI units, and/or $C_{\rm Bol}=88.697~425~...$ or if $L$ is in cgs units \citep[see IAU 2015 General Assembly Resolution B2 and][]{Eker21}. The only requirement is that the $M_{\rm Bol}$ of the star must be known beforehand then $L$ can be extracted. $M_{\rm Bol}$ is available according to following relation:
\begin{equation}
M_{\rm Bol}= M_{\rm Filter}+BC_{\rm Filter}=M_{\rm V}+BC_{\rm V},
\end{equation} 
where $M_{\rm V}$ and $BC_{\rm V}$ are the absolute visual magnitude and associated bolometric correction for the star in question. Notice that the absolute bolometric magnitude of a star is independent of the filter used in observations. Since $V$ filter observations are the oldest and most available, the $V$ filter was chosen here to symbolize all {\it BC} values of many photometric bands. If the $BC_{\rm V}$ value of the star is available, then an additional step must be taken to obtain $M_{\rm V}$ directly from observable parameters according to the following relation,

\begin{equation}
M_{\rm V}= V+5 \log \varpi + 5 - A_{\rm V},
\end{equation} 
where $V$ is the apparent visual magnitude, $\varpi$ is the trigonometric parallax of the star, which is nowadays available up to 21st magnitude \citep{Gaia20} and $A_{\rm V}$ is the extinction in the $V$ band, which can be ignorable if the star is in the local bubble \citep{Leroy93, Lallement19} or could be estimated using galactic dust maps \citep[e.g.][]{
Schlafly11, Green19}.

It is clear in this method that the only uncertainty to propagate up to $L$ comes from $M_{\rm Bol}$. It can be considered that a well-defined constant (IAU 2015 GAR B2) $C_{\rm Bol}$ does not make any contribution, thus:

\begin{equation}
{\frac{\Delta L}{L}}={\frac{\Delta M_{\rm Bol}}{2.5\log e}=0.921\times\Delta M_{\rm Bol}}.
\end{equation}

On the other hand, equation (5) indicates

\begin{equation}
\Delta M_{\rm Bol} = \sqrt{\Delta M_{\rm V}^2+\Delta BC_{\rm V}^2+ ZPE_{\rm V}^2}.
\end{equation}
which means that there are three possible error contributions to $\Delta M_{\rm Bol}$. These are: 1) random observational errors associated with the absolute visual magnitude ($\Delta M_{\rm V}$), 2) the error of the {\it BC} value itself ($\Delta BC_{\rm V}$), and 3) the zero point uncertainty of the {\it BC} scale ($ZPE_{\rm V}$). This study primarily aims to estimate the amount of the zero point error of the {\it BC} scale if the {\it BC} value comes from non-standard bolometric corrections, which are tabulated or any other source. Consequently, we may assume, just for now, that the first two contributions are zero. Then, the equation (7) changes to:

\begin{equation}
{\frac{\Delta L}{L}}=0.921\times ZPE_{\rm V}.
\end{equation}
from which one could obtain relative uncertainty of $L$ caused by the uncertainty of the zero point of the {\it BC} scale alone; that is, if the absolute visual magnitude and bolometric correction are errorless. Unfortunately, this is not the case nowadays because there are many {\it BC} sources giving non-standard {\it BC} values. If one of them is used, it is better not to omit $ZPE_V$ in equation (8). Only if one uses a standard {\it BC} in equation (5), could $ZPE_V$ in equation (8) be omitted, which corresponds to $ZPE_V$ being equal to zero, then, there is no need for equation (9). Now the question is how standard and non-standard {\it BC} values can be recognized. This is explained in the next section.

\section{Definition and Recognition of standard stellar luminosities}
\subsection{Definition of standard luminosities}
It is assured that the fixed zero point constant in equation (4) does not cause any uncertainty. Thus, the relative uncertainty $\Delta L/L$ in equation (7), should not include a term implying an uncertainty coming from $C_{\rm Bol}$ because the derivative of a constant is zero by definition.

If and only if the $BC_{\rm V}$ value used in equation (5) comes from a standard source of {\it BC} coefficients, it is not necessary to include $ZPE_V$ in equation (8). Then, it converts to
\begin{equation}
\Delta M_{\rm Bol} = \sqrt{\Delta M_{\rm V}^2+\Delta BC_{\rm V}^2}.
\end{equation}
which means there are only two error contributions to $\Delta L/L$; the observational random errors of absolute visual magnitude and the error of the {\it BC} value itself. Consequently, the $L$ obtained by equation (4), righteously, would be called standard luminosity. The concept of standard {\it BC} was originally is suggested by \citet{Eker21}.    

The term ``standard'' or ``non-standard'' before the word ``luminosity'', would be meaningful if the $L$ of any star is calculated by the method requiring a bolometric correction coefficient. If a standard {\it BC} which is already defined by \citet{Eker21} is used, the computed $L$ is called standard. To the contrary, a non-standard {\it BC} value makes the computed $L$ non-standard. Standardization of {\it BC} values, therefore, is equivalent to the standardization of stellar luminosities. According to the definition of \citet{Eker21}, the tabulated values of $BC_{\rm V}$ could be considered as standard sources for the $BC_{\rm V}$ values if the nominal value of solar absolute bolometric magnitude $M_{\rm Bol, \odot}=4.74$ mag, and the nominal solar luminosity $L_{\odot}=3.828\times10^{26}$ W as used in:  

\begin{equation}
M_{\rm Bol} = M_{\rm Bol, \odot}-2.5\log{\frac{L}{L_{\odot}}},
\end{equation}
when computing $M_{\rm Bol}$, where $L$ requires observational $R$ and $T_{\rm eff}$ from DDEB stars \citep{Eker14, Eker18}. Then, a standard $BC_{\rm V}$ is obtained according to the basic definition of bolometric correction $BC_{\rm V}=M_{\rm Bol}-M_{\rm V}$. 
\subsection{Recognizing Non-standard Luminosities}
The equations (4) and (11) are both valid for calculating $L$ of a star from its absolute bolometric magnitude. The validity is assured by:
\begin{equation}
C_{\rm Bol} = M_{\rm Bol, \odot}+2.5\log L_{\odot},
\end{equation}
according to \cite{Eker21}. Since equation (4) and the nominal values $M_{\rm Bol, \odot}=4.74$ mag, and $L_\odot=3.828\times10^{26}$ W were introduced only recently, any other non-standard values of $M_{\rm Bol, \odot}$ and $L_{\odot}$ which were used in computing a {\it BC} source (tabulated {\it BC} values or $BC_{\rm V} - T_{\rm eff}$ relation) most likely would not produce a $C_{\rm Bol}=71.197~425~...$ mag if $L$ is in SI units, and/or $C_{\rm Bol}=88.697~425~...$ mag if $L$ is in cgs units according to equation (12). This is the first and clear indication that there is a zero point error in the {\it BC} value used, which certifies that it is not a standard {\it BC} .  

It is unlikely but still under a possibility that the non-standard $M_{\rm Bol, \odot}$ and $L_{\odot}$ values would produce a $C_{\rm Bol}=71.197~425~...$ mag if $L$ is in SI units, and/or $C_{\rm Bol}=88.697~425~...$ mag if $L$ is in cgs units according to equation (12). This mathematical possibility is unavoidable because there could be an infinite number of $M_{\rm Bol, \odot}$, $L_{\odot}$ pairs to produce the same $C_{\rm Bol}$. This is the second type (an unseen indication) of a zero point error in {\it BC} values used. How to treat these two different types of zero point errors will be discussed later in the Discussion section. 

For now, let us review some of the contemporary $BC_{\rm V}$ sources in the near past with different $C_{\rm Bol}$ and corresponding non-standard $M_{\rm Bol, \odot}$ and $L_{\odot}$ values. \citet{Bessell98} gave a table for comparing the estimated $BC_{\rm V}$ of the Sun by various authors, who preferred adopting $M_{\rm Bol, \odot}$ rather than adopting $BC_{\rm V}$ of the Sun. Relying on the observed apparent visual magnitude of the Sun \citep[-26.76 mag][]{Bessell98, Torres10}, \citep[-26.74 mag][]{Allen76, Schmidt82}, \citep[-26.79 mag][]{Durrant81} and adopting either one of the quantities $M_{\rm Bol, \odot}$ or $BC_{\rm V, \odot}$ was inevitable in those years because the zero point of the $BC_{\rm V}$ scale was not yet fixed, thus it was assumed arbitrary. Adopting one of these quantities meant defining a zero point for tabulated $BC_{\rm V}$ values. Here we have reconstructed a similar table (Table 1), which enables us to compare various $C_{\rm Bol}$ values as well as the $M_{\rm Bol, \odot}$ and $L_{\odot}$ values defining it.    

% Table 1
\begin{table*}
  \centering
  \caption{Comparing $BC_{\rm V}$ values attributed to the Sun and zero point constants of bolometric magnitude scale ($C_{\rm Bol}$) and corresponding $M_{\rm Bol,\odot}$  and $L_{\odot}$ according to various authors in the near past.}
    \begin{tabular}{cccccccccl}
\hline
& $V$ & $M_{\rm V}$ & $M_{\rm Bol}$ & $BC_{\rm v}$ & $f(\times10^{6})$ & $F(\times10^{10})$ & $L(\times10^{33})$ & $C_{\rm Bol}$ & Reference\\
Order     & (mag)  & (mag) & (mag) & (mag) & (erg~cm$^{-2}\rm s^{-1}$) & (erg~cm$^{-2}\rm s^{-1}$) & (erg~s$^{-1}$) & (mag)   &\\
\hline
    1     & -26.74 & 4.83  & 4.75  & -0.08 & 1.36  & 6.284 & 3.826 & 88.70686 & \citet{Allen76} \\
    2     & -26.79 & 4.87  & 4.74  & -0.13 & 1.37  & 6.329 & 3.853 & 88.70450 & \citet{Durrant81}\\
    3     & -26.74 & 4.83  & 4.64  & -0.19 & 1.37  & 6.330 & 3.850 & 88.60365 & \citet{Schmidt82}\\
    4     & -26.76 & 4.81  & 4.74  & -0.07 & 1.371 & 6.334 & 3.856 & 88.70534 & \citet{Bessell98}\\
    5     & -26.75 & 4.82  & 4.74  & -0.08 & 1.367 & 6.322 & 3.845 & 88.70224 & \citet{Cox00}\\
    6     & -26.76 & 4.81  & 4.75  & -0.06 & 1.368 & 6.324 & 3.846 & 88.71252 & \citet{Torres10}\\
    7     & -26.76 & 4.81  & 4.75  & -0.06 & 1.361 & 6.294 & 3.828 & 88.70729 & \citet{Casagrande18}\\
    8     & -26.76 & 4.81  & 4.74  & -0.07 & 1.361 & 6.294 & 3.828 & 88.69743 & \citet{Eker20}\\
    9     & -26.76 & 4.81  & 4.645 & -0.165& 1.361 & 6.294 & 3.828 & 88.60229 & This study\\
    \hline
    \end{tabular}%
\end{table*}%

The columns of Table 1 are self-explanatory. The sequence number, apparent visual (a measured quantity), absolute visual and the adopted absolute bolometric magnitudes of the Sun are given in the first four columns. The $BC_{\rm V}$ of the Sun as the difference between absolute bolometric and visual magnitudes is in the fifth column. Corresponding solar fluxes just outside the Earth's atmosphere and on the solar surface are given in columns six and seven, while the corresponding solar luminosity is in column eight. Column nine displays the standard (row number 8) and other figures non-standard zero point constants ($C_{\rm Bol}$) according to various $BC_{\rm V}$ sources. The references are in column ten in chronological order.  

Figure 1 compares the nominated bolometric zero point constant and the nominal $M_{\rm Bol, \odot}$ or $BC_{\rm V, \odot}$ values of the Sun to the corresponding values of non-standard sources. A horizontal solid line on Figure 1a marks the nominated zero point constant ($C_{\rm Bol}=88.697~425~~...$ mag) of IAU 2015 GAR B2. 

%FIGURE 1
\begin{figure*}[ht]
\begin{center}
\includegraphics[width=\textwidth]{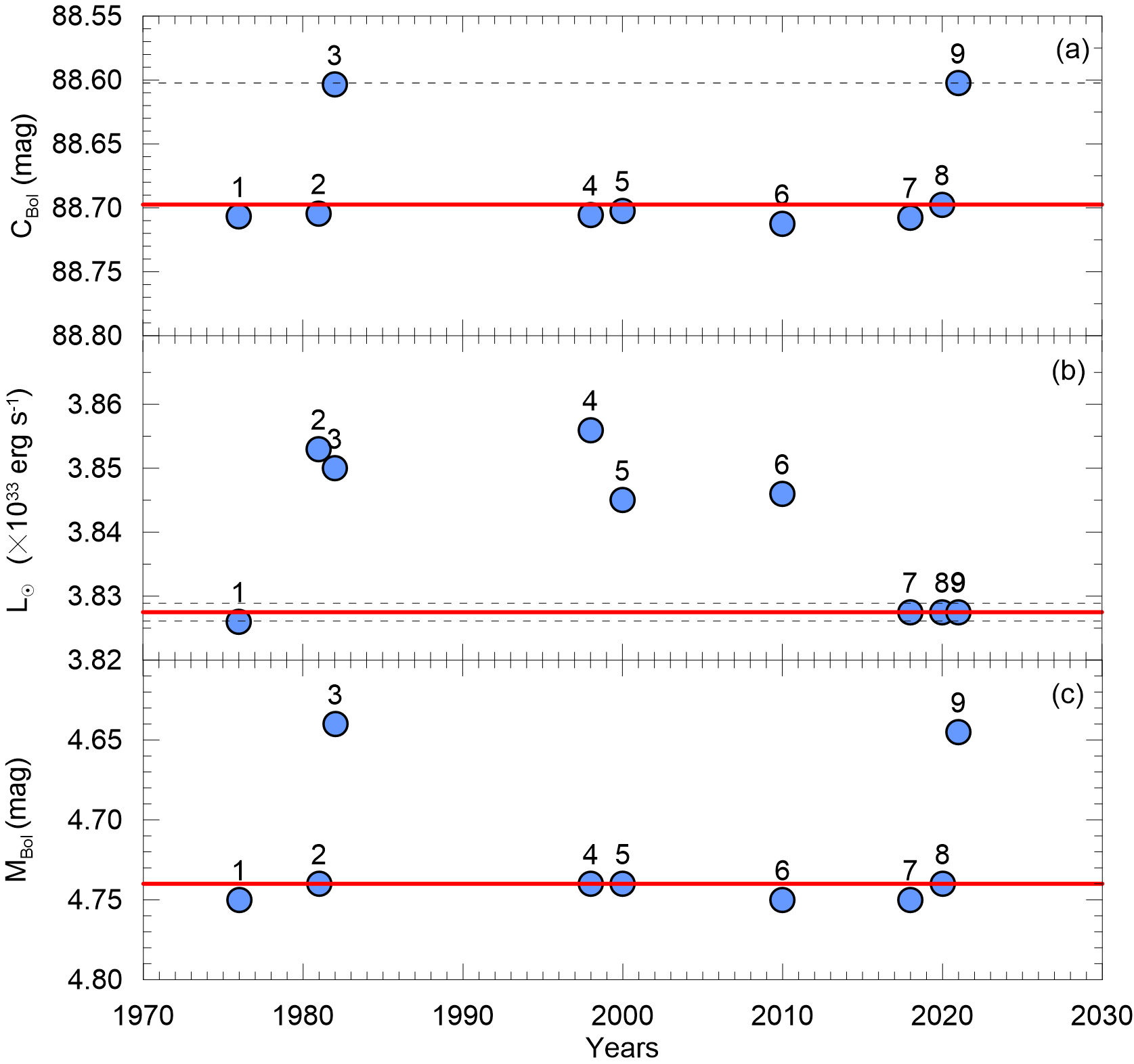}
\caption[] {Comparing nominated bolometric zero point constant and nominal $M_{\rm Bol, \odot}$ and $L_{\odot}$ values to the corresponding values of non-standard sources: 1) \citet{Allen76}, 2) \citet{Durrant81}, 3) \citet{Schmidt82}, 4) \citet{Bessell98}, 5) \citet{Cox00}, 6) \citet{Torres10}, 7) \citet{Casagrande18}, 8) \citet{Eker20}, 9) In order to shift $BC_{\rm V}-T_{\rm eff}$ curve to make all $BC_{\rm V}\leq 0$ mag (This study).}
\end{center}
\end{figure*}

Except for (3) \citep{Schmidt82} and (9), which is the test point of this study on Figure 1a with a special $M_{\rm Bol, \odot}$ (4.645 mag) to make all {\it BC} values less than zero, the other $C_{\rm Bol}$ values in the uppermost box (a) lie close to the horizontal solid line with various values which are all greater than $C_{\rm Bol}=88.697~425~...$ mag (see Table 1). The largest $C_{\rm Bol}$ (88.712523) is with \citet{Torres10}. According to Table 1 and Figure 1a, the difference between the largest and smallest $C_{\rm Bol}$ is 0.110 mag, which is equivalent to 10.13\% uncertainty on the predicted stellar luminosities even if $M_{\rm V}$ and $BC_{\rm V}$ are errorless. The least-deviated $C_{\rm Bol}$ is associated with \citet{Cox00}, which has a $C_{\rm Bol}$ value 0.005 mag bigger than the nominal value, corresponding to a 0.46\% error on $L$.                  

Figure 1b compares the nominal value of $L_{\odot}$ (horizontal solid line) to the adopted $L_{\odot}$ values (data) of the other $BC_{\rm V}$ sources. The two horizontal dashed lines marks the random observational error limits associated with the nominal solar luminosity ($L_{\odot}=3.8275\pm 0.0014\times 10^{33}$ erg s$^{-1}$) (IAU General Assembly Resolution B3). This means that in an ideal case (no error contributions from $M_{\rm V}$ and $BC_{\rm V}$ and $ZPE$), a standard stellar luminosity could be as accurate as 0.036\% (4 out of 10000). 

Please notice that, using a non-standard $L_{\odot}$ in equations (4) and (11) would produce a systematic error on $L$ even if $M_{\rm V}$ and $BC_{\rm V}$ are errorless and no zero point error exists. The largest of such systematic errors appear as an overestimation of stellar luminosities of about 0.74\% if one uses the $BC_{\rm V}$ values of \citet{Bessell98}, who has the biggest $L_{\odot}$ in Table 1. The zero point errors caused only by non-standard $L_{\odot}$ are apparently are less than 0.74\% according to Table 1 and Figure 1b. The value of $L_{\odot}$ by \citet{Cox00} in Table 1 implies a 0.46\% zero point error on Figure 1b, which confirms the same amount of error caused by his non-standard $C_{\rm Bol}$ according to Figure 1a. According to Figure 1c, because \citet{Cox00} uses $M_{\rm Bol, \odot}=4.74$ mag, which is the nominal value, one assumes no zero point error contribution from it. Despite an error of approximately about 0.46\% error on $L$ according to Table 1 and Figure 1, the {\it BC} values of \citet{Cox00} all appear to have be shifted 0.095 mag towards smaller (more negative) values compared to the {\it BC} values of \citet{Eker20}. 

The 0.095 mag systematic shift, creating a difference between the $BC_{\rm V}$ (max) of \citet{Eker20} and \citet{Cox00}, on the other hand, means an error of about 8.85\% on stellar luminosities. In other words, the $BC_{\rm V}$ values of \citet{Cox00} dominates the zero point errors and propagate within $BC_{\rm V}$ systematically. Thus, anyone who uses $BC_{\rm V}$ values of \citet{Cox00} will overestimate the stellar luminosities by about 9\%, without even including the observational errors and the error of the $BC_{\rm V}$ value used.     

Figure 1c compares the nominal value 4.74 mag (horizontal solid line) to the other adopted $M_{\rm Bol, \odot}$ (data) values of the other $BC_{\rm V}$ sources. The nominal value is that used by \citet{Durrant81}, \citet{Bessell98}, \citet{Cox00} and \citet{Eker20}. The non-standard value 4.75 mag is that used by \citet{Allen76}, \citet{Torres10} and \citet{Casagrande18}. Another non-standard value $M_{\rm Bol, \odot}=4.64$ mag is used by \citet{Schmidt82}, which is very close to our non-standard trial value of this study (Table 1, Figure 1c).  

It can be concluded here that a non-standard stellar $L$ is easy to recognize. Standard $L$ could be obtained only if standard bolometric correction coefficients were used. One can recognize non-standard bolometric corrections by checking whether nominal $M_{\rm Bol, \odot}$ and nominal $L_{\odot}$ were used or not. Although the zero point of absolute bolometric magnitudes is given as $C_{\rm Bol} = M_{\rm Bol, \odot} + 2.5\log L_{\odot}$, the nominal $C_{\rm Bol}$ does not necessarily guarantee standardization of the pre-computed $BC_{\rm V}$ values. On the contrary, a non-standard $C_{\rm Bol}$ implies non-standard $BC_{\rm V}$ values. $BC_{\rm V}$ values could be considered standard if and only if nominal $M_{\rm Bol, \odot}$ and nominal $L_{\odot}$ are used and if the zero point of the {\it BC} scale ($C_{\rm 2}$) is calculated as \citet{Eker21} describes. Authors such as \citet{Cox00}, who assume it is arbitrary, thus $C_{\rm 2}=0$, in order to make all $BC_{\rm V}$ less than zero \citep[see][and references therein]{Eker21}. This is the case when the zero point error of the {\it BC} scale cannot be estimated from pre-assumed $M_{\rm Bol, \odot}$ and $L_{\odot}$ but shows itself directly on the $BC_{\rm V}$ value itself.    

\section{Discussion}

Choosing a non-standard $L_{\odot}$ from Table 1 contributes very little (less than 1\%) to the uncertainty of a computed $L$. Therefore, the largest error contribution, definitely, comes from the choice of $M_{\rm Bol, \odot}$. This is because there are an infinite number of $M_{\rm Bol}$ and $M_{\rm Bol, \odot}$ pairs to indicate a single value of $L/L_{\odot}$ according to equation (11). On the process of computing $BC_{\rm V}$ values, $L/L_{\odot}$ is a quantity most probably coming from DDEB, thus it is a fixed value. Since $BC_{\rm V}=M_{\rm Bol}-M_{\rm V}$, where $M_{\rm V}$ is also a fixed value. Therefore, pre-computed $BC_{\rm V}$ values are effected directly by the choice of $M_{\rm Bol, \odot}$ with the classical method using equation (11). As in the case of \citet{Casagrande18}, who preferred to use $M_{\rm Bol, \odot}=4.75$ rather than the nominal 4.74 mag, all pre computed $BC_{\rm V}$ values would be 0.01 mag smaller compared to the standard $BC_{\rm V}$. Although \citet{Casagrande18} do not state clearly which $L_{\odot}$ value used, we have assumed they are using the nominal $L_{\odot}$ since they cite IAU 2015 GAR B3 for the value of $L_{\odot}$ used in their equation (3) for obtaining the bolometric flux received from a star. 

In another aspect, $M_{\rm Bol, \odot}$ acts as the arbitrary zero point for bolometric magnitudes, as \citet{Casagrande18} propose that ``any value is equally legitimate on the condition that once chosen, all bolometric corrections are scale accordingly''. That is, different authors using different $M_{\rm Bol, \odot}$ end up calculating different $BC_{\rm V}$ values for the same star. Because $BC_{\rm V}=M_{\rm Bol}-M_{\rm V}$, the zero point problem shows itself in the produced $BC_{\rm V}$ values. Using the nominal values of $M_{\rm Bol, \odot}$ and $L_{\odot}$ \citet{Eker20} produced a standard $BC_{\rm V}-T_{\rm eff}$ relation, which has a maximum $BC_{\rm V}$ value of $BC_{\rm V}$ (max) = 0.095 mag at $T_{\rm eff}=6897$ K. For this study we have searched for the value of $M_{\rm Bol, \odot}$, without changing the nominal value of $L_{\odot}$, that would obtain a $BC_{\rm V}-T_{\rm eff}$ relation with $BC_{\rm V}$ (max) = 0.00. We concluded that the answer is $M_{\rm Bol, \odot}=4.645$ mag; then, all computed {\it BC} values will be reduced (more negative) to be less than zero, as listed in \citet{Cox00}. However, such a relation is cannot be definitely considered standard. 

The problem with the $BC_{\rm V}$ values of \citet{Cox00} is not the same as adopting $M_{\rm Bol, \odot}=4.645$ mag in order to make all $BC_{\rm V}$ values negative. His $BC_{\rm V}$ values are all negative, despite his use of $M_{\rm Bol, \odot}=4.74$ mag. In fact, his $C_{\rm Bol}$ value is the nearest $C_{\rm Bol}$ among the other sources, which wrongly imply that if one uses a $BC_{\rm V}$ value of \citet{Cox00}, he/she would obtain a very accurate luminosity (0.46\%) despite, in reality, the computed $L$ having a 9\% systematic zero point error. Most probably, the $BC_{\rm V}$ values of \citet{Cox00} were calculated using the following definition of $BC_{\rm V}$:    

\begin{equation}
BC_{\rm V} = 2.5\log \frac{f_{\rm V}}{f_{\rm Bol}}+(C_{\rm Bol}-C_{\rm V}) = 2.5\log \left(\frac{\int_0^\infty \!S_{\lambda}(V)f_{\lambda}\rm d\lambda}{\int_0^\infty \!f_{\lambda}\rm d\lambda}\right)+C_2.  
\end{equation}
where $C_2$, the zero point constant of the {\it BC} scale, was assumed arbitrary, thus, $C_2=0$ value was taken arbitrarily. If $C_2$ is assumed zero, all $BC_{\rm V}$ values becomes unquestionably be less than zero. This is because visual flux ($f_{\rm V}$) is never zero but less than the bolometric flux ($f_{\rm Bol}$). The logarithm of numbers between zero and one ($f_{\rm V}/f_{\rm Bol}$) is always negative, which requires a positive $C_2$ otherwise (if $C_2=0$ or $C_2<0$) the {\it BC} scale does not have a zero point; a negative number plus zero or adding two negative numbers does not produce a number zero. The {\it BC} producers, who assumed that the zero point constant of the {\it BC} scale is arbitrary, believed that they had the right to make it zero. This way, they unknowingly carried the zero point error into the {\it BC} value itself.

Obviously there are two approaches to remove the zero point errors. The first approach is that suggested by \citet{Casagrande18} and \citet{Torres10}. This entails being cautious when calculating the $M_{\rm Bol}$ of a star from its $M_{\rm V}$ and $BC_{\rm V}$. Before using the $BC_{\rm V}$ on the formula $M_{\rm Bol} = M_{\rm V} + BC_{\rm V}$, check it out and first learn which $M_{\rm Bol, \odot}$ and $L_{\odot}$ values were used before producing the tabulated $BC_{\rm V}$ values or $BC_{\rm V}-T_{\rm eff}$ relation from which the $BC_{\rm V}$ value was taken. Then, it is safe to proceed in calculating $M_{\rm Bol} = M_{\rm V} + BC_{\rm V}$ in the first step. However, do not use equation (4); instead, use equation (11) when calculating $L$. Do not forget to use the same $M_{\rm Bol, \odot}$ and $L_{\odot}$ values, which you have searched for in order to ensure that they are consistent with the $BC_{\rm V}$ value you are using.     

The second approach is suggested in this study. Do not use any value of $BC_{\rm V}$ haphazardly. Use only standard $BC_{\rm V}$ values from standard sources. You can use either one of the equations (4) or (11). It does not matter which, since both are valid for producing the standard $L$ of stars. 

Notice that, the first approach fails to produce a standard $L$ if one uses $BC_{\rm V}$ values from the sources such as \citet{Cox00} which may appear to be using nominal $M_{\rm Bol}$ and $L_{\odot}$ but their $BC_{\rm V}$ values are not standard because they were produced by assuming $C_2=0$. Concentrating only on the difference between the $M_{\rm Bol}$ of a star and $M_{\rm Bol}$ of the Sun, the first approach has no answer as to which $L_{\odot}$ was used if the stellar $L$ is taken from published papers where the published $L$ are usually expressed in solar units.    

There are no such problems in the second approach.

\subsection{Typical Accuracy of L in Method 1 (direct method)}
The  peak of the histogram distribution of relative radius errors of DDEB is 2\% according to \citet{Eker14}. For today's accuracy we may take it as 1\%, since many newly-published papers give stellar radii even more accurate than $\sim 1\%$. According to \citet{Eker15}, typical temperature accuracy is 2-3\%. The acceptable stellar effective temperature uncertainty for single stars in general is 1-2.5\% according to \citet*{Masana06}. Since the direct method uses the effective temperatures of DDEB, we may find a typical temperature uncertainty of 2-3\%. Consequently, equation (1) indicates that the typical uncertainty range of stellar $L$ is 8.2\% - 12.2\%. On the more accurate side, there could be stars like the primary component of V505 Per having $\Delta R/R = 1.09\%$ and $\Delta T_{\rm eff}/T_{\rm eff} = 0.32\%$ with corresponding $\Delta L/L = 2.53\%$ \citep{Tomasella08}. That is, the accuracy of a few percentage levels has been already achieved by the direct method.

\subsection{Typical Accuracy of L in Method 2 (using MLR)}
Among the three methods, the method using a MLR is the least accurate. This is because the relative uncertainty of $L (\Delta L/L)$ is determined by the standard deviation ($SD$) of stellar luminosities on a $\log M-\log L$ diagram according to equation (3), where $SD$ could be different at different mass ranges \citep{Eker15, Eker18}. The most accurate mass range, which was named the ultra-low mass domain covering stellar masses in the range $0.179 < M/M_{\odot} < 0.45$, has $SD =0.076$ mag which corresponds to $\Delta L/L = 17.5\%$. The most dispersed range, which is called the high mass domain covering stellar masses in the range $2.4<M/M_{\odot}\leq 7$, has $SD = 0.165$ mag which corresponds to $\Delta L/L=37.99\%$ \citep{Eker18}.

\subsection{Typical Accuracy of L in Method 3}
\subsubsection{Typical Accuracy of L using a standard {\it BC}}
Standard stellar luminosities are those calculated by the method using a pre-determined standard $BC_{\rm V}$. The most important problem with this method is that the systematic zero point errors of non-standard $BC_{\rm V}$ could be as big as 0.11 mag, corresponding to 10.13\% errors in the predicted stellar luminosities. Such systematic errors could be removed if and only if one uses standard $BC_{\rm V}$ sources. To make the comparison meaningful, here we assume the zero point error has already been removed or taken to be zero.

A typical accuracy of standard $L$, therefore, could be calculated by equations (7) and (8) but taking $ZPE_{\rm V}$ as zero in the equation (8).

There could be three contributions to the uncertainty of $\Delta M_V$ according to equation (6), \citet{Eker20} gives:

\begin{equation}
\Delta M_V = \sqrt{\left(\Delta m_V \right)^2+\left(5\log e\frac{\sigma_\varpi}{\varpi}\right)^2+ \left(\Delta A_V \right)^2,}
\end{equation} 
where, the first term in the square root stands for the error contribution from the apparent brightness, the second term represents the contribution from stellar parallaxes, and the last term is for the contribution from interstellar extinctions. Nowadays, apparent magnitude uncertainties are in the order of milli magnitudes. If we assume extinction errors ($\Delta A_{\rm V}$) are about 0.01 mag, definitely the parallax errors would definitely dominate the others; If there is a 10\% error on the parallax, $\Delta M_{\rm V}=0.217$ mag. For stars with a 5\% error in their parallaxes, $\Delta M_{\rm V}=0.109$ mag. Histogram distribution of the parallaxes for 206 DDEB (400 stars), from which the standard $BC_{\rm V}$ was extracted \citep{Eker20}, has a peak (median) of 2\%. If we take this as a typical parallax error, then typically $\Delta M_{\rm V}=0.044$ mag.       

The typical standard error of a $BC_{\rm V}$ value, if it comes from a standard $BC_{\rm V}-T_{\rm eff}$ curve, would be in the orders of 0.011 mag (Table 2) for the range of temperatures (3000-36000 K) considered for main-sequence stars by \citet{Eker20}. The re-arranged data indicates a standard error of $BC_{\rm V}-T_{\rm eff}$ being reduced to 0.009 mag for medium temperatures (5000-10000 K) while it is 0.59 mag for lower temperatures, and 0.028 mag for the higher temperatures. Inserting typical $\Delta M_{\rm V}=0.044$ mag and a typical standard error of a {\it BC} (0.009 mag) into equation (10), a typical $\Delta M_{\rm Bol}=0.045$ mag is obtained for the middle temperatures. Inserting this into equation (7), a typical $\Delta L/L$ becomes 4.14\%. 

% Table 2
\begin{table}
  \centering
  \caption{Standard deviation of data from $BC_{\rm V}-T_{\rm eff}$ curve of \citet{Eker20} and standard errors of $BC_{\rm V}$ values.}
    \begin{tabular}{cccc}
    \hline
$T_{\rm eff}$ range (K) & $N$   & $SD$ & $SD/\sqrt{N}$\\
\hline
    3000-5000  &  35  & 0.349 & 0.059 \\
    5000-10000 & 261  & 0.142 & 0.009 \\
    10000-36000& 104  & 0.290 & 0.028 \\
    3000-36000 & 400  & 0.215 & 0.011 \\
    \hline
    \end{tabular}%
  \label{tab:addlabel}%
\end{table}%

Now, let us assume the extreme case of a star (CM Dra) with a relative parallax error of 0.050\% \citep{Gaia20}. We may assume no extinction because it is only 14.86 pc away according to {\it Gaia} eDR3 data. Then, the milli-magnitude accuracy of its apparent brightness would imply $\Delta M_{\rm V}=0.0019$ mag. Assuming its $BC_{\rm V}$ has a standard error of 0.011 mag ($4^{\rm th}$ line in Table 2), its standard luminosity would have an uncertainty of 1.03\%. Accuracy in standard luminosity even increases to 0.82\% for the middle temperatures. 

Here, we run into an unexpected case: the secondary method, using a {\it BC}, provides more accurate stellar luminosities than the direct method. It is too good to be true. Apparently, the problem must be taking the standard error of a standard $BC_{\rm V}-T_{\rm eff}$ curve as the standard error of a the {\it BC} value before computing $\Delta M_{\rm Bol}$. 

As \citet{Eker21} stated, a standard $BC_{\rm V}-T_{\rm eff}$ curve obtained from $BC_{\rm V}$ and $T_{\rm eff}$ values is similar to mass-luminosity relations obtained from masses and luminosities, but not like the Planck curve representing the spectral energy distribution (SED) of a star. Thus, the standard error of a mass-luminosity relation (curve) cannot be used as the standard error of an $L$ for a given $M$. This is because the scattering of data from the relation is not only due to observational errors but also due to the different ages and chemical compositions of the stars on the $M-L$ diagram. Therefore, $\Delta L/L$ should be calculated directly from the standard deviations, as described in Section 2.2 

Similarly, one must not use standard errors (column 3 of Table 2) but use the standard deviations (column 2 of Table 2) when computing typical errors of $M_{\rm Bol}$. If this is done, 0.215 mag standard deviation for the total range indicates $\Delta L/L=0.198$, and 0.142 mag standard deviation for the middle temperatures indicates $\Delta L/L=0.131$ for the star CM Dra which has $\Delta M_{\rm V}=0.0019$ mag. For a typical $\Delta M_{\rm V}=0.044$ mag, $\Delta L/L$ becomes 20.2\% for all main-sequence temperatures in general, and 13.7\% for the middle temperature only. Therefore, the indirect method of using standard {\it BC} when computing standard stellar luminosities could be considered better than using MLR, but not as good as the direct method. 

\subsubsection{Typical Accuracy of $L$ using a Unique {\it BC}}
Another advantage of the method of computing the standard $L$ of a star using its {\it BC} is when one does not need a pre-computed $BC_{\rm V}-T_{\rm eff}$ curve or a table giving standard {\it BC} values. The unique {\it BC} of a star may be obtained directly from its SED according to equation (13). Using this equation, however, requires spectroscopic observation of stars covering a sufficient spectral range at least more than the photometric filter which is used in determining its absolute magnitude in equation (6). Interstellar reddening is still there to spoil SED, but restoring of the SED is always possible as done by \citet{Stassun16}. This may even not be necessary if the star is in the local bubble where interstellar extinction could be ignored. 

Assuming that $C_2$ is a well-defined quantity \citep{Eker21}, the error of {\it BC} in this method depends on how accurate $f_{\rm V}/f_{\rm Bol}$ could be determined. If the quantity $f_{\rm V}/f_{\rm Bol}$ is determined in a few percentage level, since there is no error contribution from $C_2$, then according to equation (5) the accuracy of $M_{\rm Bol}$ depends on the accuracy of two quantities, $M_{\rm V}$ and $BC_{\rm V}$, which means that the $M_{\rm Bol}$ of a star could be obtained at a low percentage. For example, using typical $\Delta M_{\rm V}=0.044$ mag as pointed out in Section 5.3.1, assuming $\Delta BC_{\rm V}=0.03$ mag, $\Delta M_{\rm Bol}$ is 0.053 mag which means $\Delta L/L$ is 4.9\%. If a star has a very accurate parallax like CM Dra, then its absolute visual magnitude could be as accurate as 0.0019 mag, and if its {\it BC} is calculated with an accuracy of 1\%, then its luminosity would be 0.9\%. 

We estimate here that the typical accuracy of a standard stellar luminosity could be in the order of a few percent levels if the method uses the star's unique {\it BC}, which could be computed from its SED. For ideal cases it could be as accurate as 1\% or even more, depending upon the accuracies of $M_V$ and the unique {\it BC} itself. Therefore, using a unique {\it BC} which is computed from its SED is a real advantage over all other methods including the direct method. 

\section{Conclusions}
The accuracy of predicted stellar luminosities using the direct and two secondary methods has been revised and a new concept, ``standard stellar luminosity'', has been defined. The luminosities which were calculated by the direct method from observational radii and effective temperatures are more accurate than the luminosities estimated by the secondary methods which require a MLR or a pre-computed standard {\it BC}. The luminosities produced by the direct method have been shown to be accurate from about 8.2\% to 12.2\%, and they could be more accurate by 2-3\%. Stellar luminosities obtained by the direct method are standard by definition because there could be only one luminosity for a star of given radius and effective temperature.

The stellar luminosities by the method using MLR are the least accurate. Depending upon the mass range where the classical MLR operates in the form $L\propto M^{\alpha}$, they could be most accurate at about 17.5\% (for low mass stars; $0.179\leq M/M_{\odot}\leq 0.45$) and least accurate at about 38\% (for high mass stars; $2.4\leq M/M_{\odot}\leq 7$). Luminosities calculated by a method with MLR cannot be called standard because, first of all, a standard MLR has not been defined yet, and perhaps it never will. This is because a MLR in the form $L\propto M^{\alpha}$ is the relation between the typical mass and typical luminosity of a main-sequence star belonging to a certain region (e.g., Galactic disc stars in the solar neighborhood). Moreover, there is no relation or process providing the true luminosity of a main-sequence star from its mass except a standard stellar structure and evolutionary model (not yet defined or agreed upon), which naturally would include the chemical composition and age, as with the other independent parameters.

The stellar luminosities produced by the method using a pre-determined {\it BC} are more accurate than the luminosities computed by the method of using MLR, but less accurate than the luminosities produced by the direct method. It has been shown that the method which uses a standard {\it BC} would provide 13.7\% accuracy in predicted luminosities if the typical accuracy of the absolute visual magnitude is about $\pm0.044$ mag for main-sequence stars with middle temperatures of $5000 \leq T_{\rm eff}\leq 10000$ K. Depending upon other factors, such as the accuracy of its parallax, interstellar extinction, and accuracy of the standard {\it BC}, the accuracy would be improved or becomes worse. It has been said that such luminosities are called standard only if the {\it BC} value used in computing it is a standard {\it BC}. Non-standard {\it BCs} produce non-standard luminosities.   

Note that if a {\it BC} value is from a tabulated standard {\it BC} table and/or from a standard $BC_{\rm V}-T_{\rm eff}$ relation, the {\it BC} value cannot be called unique. Tabulated standard {\it BC} tables and/or $BC_{\rm V}-T_{\rm eff}$ relations in the literature do not provide unique {\it BC} values because they have already been produced from pre-computed unique {\it BC} values; thus, they provide only a mean value to represent stars of a given effective temperature at different age and chemical composition.

Therefore, a unique {\it BC} is a {\it BC} computed according to equation (13) for a star from its spectrum with sufficient spectral coverage and resolution. Unlike tabulated tables and/or $BC_{\rm V}-T_{\rm eff}$ relations where chemical composition and age information has been lost, a computed $f_{\rm V}/f_{\rm Bol}$ from the well-observed spectrum of a star retains this information;  therefore, such computed {\it BC} values of stars are unique. Using unique {\it BC} when computing standard luminosity has the advantage of providing an even more accurate standard $L$ than the other methods, including the direct method. 

Unfortunately, this new method, described in the present study, has no application yet because it requires $C_2$ (equation 13) for the $V$ band and other bands, which are not available in the literature \citep{Eker21}. Therefore, we encourage determining $C_2$ values for various bands first, and then the determinations of $f_{\rm V}/f_{\rm Bol}$ from observed stellar spectra in order to compute a unique $BC_{\rm V}$ value for each star. Then, unique stellar luminosities could be computed using equations (4), (5) and (6).

\section*{Acknowledgements}
We are grateful to anonymous referee whose comments was useful improving the revised manuscript. Thanks to Mr. Graham H. Lee for careful proofreading of the text and correcting its English grammar and linguistics. This work has been supported in part by the Scientific and Technological Research Council (T\"UB\.ITAK) under the grant number 114R072. Thanks are also given to the BAP office of Akdeniz University for providing partial support for this research.

\section*{Data Availability}
The data underlying this article are available at https://dx.doi.org/10.1093/mnras/staa1659



%\bibliographystyle{mnras}

% Don't change these lines
\bsp	% typesetting comment
\label{lastpage}
\end{document}